\documentstyle[12pt]{article}

\begin{document}

\def\be{\begin{equation}}
\def\ee{\end{equation}}
\def\bea{\begin{eqnarray}}
\def\eea{\end{eqnarray}}
\def\nn{\nonumber \\}
\def\e{{\rm e}}

\  \hfill
\begin{minipage}{3.5cm}
November 2002 \\
\end{minipage}

\vfill

\begin{center}
{\large\bf Universal features of the holographic duality:
conformal anomaly and brane gravity trapping from 5d AdS Black Hole}

\vfill

{\sc Shin'ichi NOJIRI}\footnote{nojiri@cc.nda.ac.jp}
and {\sc Serguei D. ODINTSOV}$^{\spadesuit}
$\footnote{
odintsov@mail.tomsknet.ru}

\vfill

{\sl Department of Applied Physics \\
National Defence Academy,
Hashirimizu Yokosuka 239-8686, JAPAN}

\vfill

{\sl $\spadesuit$
Lab. for Fundamental Studies,
Tomsk State Pedagogical University,
634041 Tomsk, RUSSIA}

\vfill

{\bf ABSTRACT}

\end{center}

We calculate the holographic conformal anomaly and brane Newton potential 
when bulk is 5d AdS BH. It is shown that such anomaly is the same as 
in the case of pure AdS or (asymptotically) dS bulk spaces, i.e. it is 
(bulk) metric independent one. While Newton potential on the static brane 
in AdS BH is different from the one in pure AdS space, the gravity trapping 
still occurs for two branes system. This indicates to metric independence of 
gravity localization.

\vfill


\newpage

\section{Introduction}

The spectacular realization of the holographic principle in string theory in
the form of (A)dS/CFT correspondence and braneworld scenario shows that
there are some close 
relations between so far distant areas of the theory of fundamental
interactions.
Moreover, some really new questions should be addressed in the connection 
with holography.
Indeed, the first impression of the AdS/CFT correspondence is the
fundamental role of bulk AdS space. On the same time, it became clear that 
different bulk spaces (say,  pure AdS or AdS Black Hole or even  
dS) simply correspond to different dual CFTs when duality may be established.
In this connection, the fundamental question may be: what are the universal 
features of holographic duality? In other words, what properties do not
depend on the choice of bulk space?

In the present contribution we discuss two issues: the calculation
of the 4d holographic conformal anomaly and of the brane Newton potential
when bulk space is 5d AdS BH. It is shown that the holographic conformal
anomaly has the same form as in case of pure AdS or pure dS bulk space. 
This is demonstration of its bulk metric (as well as horizon) independence.
The Newton potential on the (adiabatically) static brane (when bulk is AdS
BH) is different from the one in case of pure AdS bulk. Nevertheless, 
for large BH limit the gravity trapping occurs for two branes model.
This indicates that gravity trapping is also universal phenomenon which
does not depend from the choice of bulk space.  

\section{ Metric independence of holographic conformal anomaly: 
bulk black hole spacetime}

In this section our purpose is to calculate holographic conformal 
anomaly from bulk AdS black hole, using AdS/CFT correspondence\cite{AdS}.
We prove that such holographic anomaly turns out to be the same as from
pure bulk AdS space. The same phenomenon occurs for 5d cosmological 
deSitter bulk space. This explicitly shows  (bulk)  metric independence as
well as 
horizon independence of 
holographic conformal anomaly.

One starts from 
5-dimensional AdS-Schwarzschild black hole background:
\be
\label{AdSS}
ds_{\rm AdS-S}^2 = -\left({r^2 \over l^2} - {\mu \over r^2}
\right) dt^2 +  \left({r^2 \over l^2} - {\mu \over r^2}
\right)^{-1} dr^2 + r^2 \sum_{i=1}^3\left(dx^i\right)^2\ .
\ee
The last term in (\ref{AdSS}) is replaced as 
\be
\label{S1}
ds^2 = -\left({r^2 \over l^2} - {\mu \over r^2}
\right) dt^2 +  \left({r^2 \over l^2} - {\mu \over r^2}
\right)^{-1} dr^2 + r^2 \sum_{i=1}^3g_{ij}dx^i dx^j\ ,
\ee
and we introduce a new coordinate $\rho$ by
\be
\label{S2}
\rho\equiv r^{-2}\ ,
\ee 
and rewrite the metric (\ref{S1}) in the following form:
\be
\label{S3}
ds^2 = -\left({1 \over l^2\rho} - \mu \rho\right) dt^2 
+ {1 \over 4}\left({\rho^2 \over l^2} - \mu\rho^4\right)^{-1} 
d\rho^2 + \rho^{-1} \sum_{i,j=1}^3g_{ij}dx^i dx^j\ ,
\ee
In the limit $\rho\to 0$ ($r\rightarrow \infty$), the metric 
behaves as
\be
\label{S4}
ds^2 \to {l^2 \over 4\rho^2} d\rho^2 
+ \rho^{-1}\left(-{1 \over l^2} 
+ \sum_{i=1}^3g_{ij}dx^i dx^j\right)
= {l^2 \over 4\rho^2} d\rho^2 
+ \rho^{-1}\sum_{m,n=0}^3 \tilde g_{mn}dx^m dx^n\ .
\ee
Then one can regard that for the general 
metric $\tilde g_{mn}$, we have chosen gauge conditions:
\be
\label{S5}
\tilde g_{tt}=-{1 \over l^2}\ ,\quad \tilde g_{ti} 
=\tilde g_{it}=0\ ,(i=1,2,3)\ .
\ee
For the metric $\tilde g_{mn}$, the non-vanishing components 
of the connecetion $\tilde \Gamma^l_{mn}$ and the Ricci curvature 
$\tilde R_{mn}$ are 
\bea
\label{S6}
&& \tilde \Gamma^t_{ij} = {l^2 \over 2}g_{ij,t}\ ,\quad 
\tilde \Gamma^i_{tj} = \tilde \Gamma^i_{jt} 
= {1 \over 2}g^{ik}g_{kj,t}\ ,\quad 
\tilde \Gamma^i_{jk}=\gamma^i_{jk}\ ,\nn
&& \tilde R_{tt} = - {1 \over 2}g^{ij}g_{ij,tt} 
+ {1 \over 4}g^{ik}g^{jl}g_{ij,t}g_{kl,t} \ ,\nn
&& \tilde R_{ij} = r_{ij} + {l^2 \over 2}g_{ij,tt} 
 - {l^2 \over 2}g^{kl}g_{ki,t}g_{lj,t} 
 + {l^2 \over 4}g_{ij,t}g^{kl}g_{kl,t} \ ,\nn
&& \tilde R = r + 2l^2 g^{ij} g_{ij,tt} 
 - {3l^2 \over 4}g^{ij}g^{kl}g_{ki,t}g_{lj,t} 
 + {l^2 \over 4}\left(g^{kl}g_{kl,t}\right)^2\ .
\eea
Here $r^i_{jk}$, $r_{ij}$ and $r$ are the connection, the 
Ricci tensor and the scalar curvature given by $g_{ij}$. 

For the metric (\ref{S3}), the non-trivial components of the 
connection are
\bea
\label{S7}
&& \Gamma^\rho_{\rho\rho}=- {1 \over \rho}
{1 -  2\mu l^2\rho^2 \over 1 - \mu l^2\rho^2}
\sim - {1 \over \rho}\left(1 - \mu l^2\rho^2 
+ {\cal O}\left(\rho^4\right)\right) \ ,\nn
&& \Gamma^\rho_{tt} = - {2 \over l^4}
\left(1 -  \mu l^2\rho^2\right)\left( 1 + \mu l^2\rho^2\right) 
\sim - {2 \over l^4}\left(1 + {\cal O}\left(\rho^4\right)\right) \ ,\nn
&& \Gamma^t_{\rho t}=\Gamma^t_{t\rho}= - {1 \over 2\rho}
{1  + \mu l^2\rho^2 \over 1 - \mu l^2\rho^2}
\sim - {1 \over 2\rho}\left(1 + 2 \mu l^2\rho^2 
+ {\cal O}\left(\rho^4\right)\right) \ ,\nn
&& \Gamma^\rho_{ij}= {2 \over l^2} \left( 1 - \mu l^2\rho^2\right)
\left(g_{ij} - \rho g_{ij,\rho}\right)\ ,\quad 
\Gamma^i_{j\rho}=\Gamma^i_{\rho j} = - {1 \over 2}\left(
{1 \over \rho} - g^{ik}g_{kj,\rho}\right)\ ,\nn
&& \Gamma^t_{ij}={l^2 \over 2}\left( 1 -  \mu l^2\rho^2
\right)g_{ij,t} \ ,
\quad \Gamma^i_{jt}=\Gamma^i_{tj}={1 \over 2}g^{ik}g_{kj,t}\ ,
\quad \Gamma^i_{jk}=\gamma^i_{jk}\ .
\eea
As we are interested in the holographic conformal anomaly, 
we calculate the curvatures up to relevant order in 
the power of $\rho$. Then the non-trivial components are 
given by 
\bea
\label{S8}
R_{\rho\rho} &=& - {1 \over \rho^2} - \mu l^2 
 - {1 \over 2}g^{ik}g_{ik,\rho\rho} 
 + {1 \over 4} g^{ik}g^{jl} g_{ij,\rho} g_{kl,\rho} 
 + {\cal O}\left(\rho\right) \ ,\nn
R_{tt} &=& {4 \over l^4 \rho} - {4\mu\rho \over l^2} 
 - {1 \over l^4}g^{ik}g_{ki,\rho} + \tilde R_{tt} 
+ {\cal O}\left(\rho^2\right)  \ ,\nn
R_{ij} &=& - {4 \over l^2\rho}g_{ij} + {2 \over l^2}g_{ij,\rho} 
 + {1 \over l^2}g_{ij} g^{kl}g_{kl,\rho} 
 - {2\rho \over l^2}g_{ij,\rho\rho} \nn
&& + {2\rho \over l^2}g^{kl}g_{ki,\rho}g_{lj,\rho} 
 - {\rho \over l^2} g_{ij,\rho}g^{kl}g_{kl,\rho} 
 + \tilde R_{ij} \ ,\nn
R &=& - {20 \over l^2} + \rho \tilde R 
+ \rho^2 \left( - {4 \over l^2}g^{ij}g_{ij,\rho\rho} 
+ {3 \over l^2}g^{ik}g^{jl}g_{ij,\rho}g_{kl,\rho} \right. \nn
&& \left. - {1 \over l^2}\left(g^{ij}g_{ij,\rho}
\right)^2 \right) + {\cal O}\left(\rho^3\right)\ .
\eea
When one calculates the holographic conformal anomaly, 
 $\tilde g_{mn}$ or $g_{ij}$ are expanded as a power serie 
of $\rho$,
\be
\label{S9}
g_{ij}=g^{(0)}_{ij} + \rho g^{(1)}_{ij} 
+ \rho^2 g^{(2)}_{ij} + \cdots\ .
\ee
By using the Einstein equation, $g^{(1)}_{ij}$, 
$g^{(2)}_{ij}$, $\cdots$ can be solved with respect 
to $g^{(0)}$. After that,  substituting these 
expressions into the Einstein-Hilbert action, one can 
find the holographic anomaly from the coefficient of $\rho^{-1}$ 
term\cite{anom}. From Eq.(\ref{S8}), the $\mu$-dependent term 
does not contribute to the Einstein equation in the 
relevant order. Since $R$ does not contain $\mu$-dependent 
term (they are cancelled with each other), the 
$\mu$-dependent term does not appear in the expression for 
the conformal anomaly, what is consistent with the usual 
field theory calculation. 

If we put $\mu=0$, the metric  (\ref{S4}) is invariant if we change $\rho$ 
and $\tilde g_{ij}$ by 
\be
\label{wtr}
\delta\rho= \delta\sigma\rho\ ,\ \ 
\delta \tilde g_{ij}= \delta\sigma g_{ij}\ .
\ee
Here $\delta\sigma$ is a constant parameter of the transformation.
The transformation (\ref{wtr}) can be regarded as the scale 
transformation. When one substitutes the expressions in (\ref{S9}) 
(after solving $g^{(1)}_{ij}$ etc. with respect to $g^{(0)}_{ij}$) into 
the action, the action diverges in general since the action 
contains the infinite volume integration on the asymptotically AdS 
space. 
The action is regularized by introducing the infrared cutoff 
$\epsilon$, which generates a boundary at finite $\rho$ 
($=\epsilon$)
\be
\label{vi}
\int d^5 x\rightarrow \int d^4 x\int_\epsilon d\rho \ ,\ \ 
\int_{{\rm Boundary}} d^4 x\Bigl(\cdots\Bigr)\rightarrow 
\int d^4 x\left.\Bigl(\cdots\Bigr)\right|_{\rho=\epsilon}\ .
\ee
The terms proportional to the (inverse) power of 
$\epsilon$ in the regularized action are invariant under the scale 
transformation 
\be
\label{via}
\delta g_{(0)\mu\nu}=2\delta\sigma g_{(0)\mu\nu}\ ,\ \  
\delta\epsilon=2\delta\sigma\epsilon \ , 
\ee
which corresponds to (\ref{wtr}).  
Then the subtraction of these terms proportional to the inverse power of 
$\epsilon$ does not break the invariance under the scale transformation. 
When $d$ is even, however, the term proportional to $\ln\epsilon$ appears.
This term is not invariant under the scale transformation (\ref{via}) and the 
subtraction of the $\ln\epsilon$ term breaks the invariance. 
The variation of the $\ln\epsilon$ term under the scale 
transformation (\ref{via}) is finite when $\epsilon\rightarrow 0$ and 
should be canceled by the variation of the finite term (which does not 
depend on $\epsilon$) in the action since the original action 
is invariant under the scale transformation. 
Therefore the $\ln\epsilon$ term $S_{\rm ln}$ gives the Weyl anomaly $T$ 
of the action renormalized by the subtraction of the terms which diverge 
when $\epsilon\rightarrow 0$ by 
\be
\label{vib}
S_{\rm ln}=-{1 \over 2}
\int d^4x \sqrt{-g_{(0)}}T\ .
\ee
The explicit form of $T$ is found to be (for explicit calculations, see
\cite{anom})
\be
\label{II11}
T={l^3 \over 8\pi G} 
\left[ {1 \over 8}R_{(0)ij}R_{(0)}^{ij}
-{1 \over 24}R_{(0)}^2 \right]\ .
\ee
Comparing with the field theory calculation, the  
conformal anomaly coming from the multiplets of ${\cal N}=4$ 
supersymmetric $U(N)$ or $SU(N)$ Yang-Mills, we obtain
\be
\label{II13}
{l^3 \over 16\pi G}={2N^2 \over (4\pi)^2}\ .
\ee
Hence, the calculation of holographic anomaly represents explicit check 
of AdS/CFT correspondence. Moreover, the previous calculations were limited
to pure (or asymptotically) AdS  spaces \cite{anom}.
From above AdS BH calculation one arrives at the conclusion of bulk metric
independence of holographic anomaly.
This is supported also by bulk deSitter space case where 
dual calculation for anomaly gives precisely above result\cite{dsanom}.

Moreover, one can also consider asymptotically deSitter space instead of
asymptotically
anti-deSitter space. 
If one replaces the length parameter $l^2$, the time coordinate $t$ and the
radial 
coordinate $r$ by
\be
\label{dS1}
l^2\to -l^2\ ,\quad t\to r\ ,\quad r\to t\ ,
\ee
in Eq.(\ref{AdSS}), we obtain the asymptotically deSitter space as 
\be
\label{dS2}
ds_{\rm dS-S}^2 = - \left({t^2 \over l^2} + {\mu \over t^2}
\right)^{-1} dt^2 + \left({t^2 \over l^2} + {\mu \over r^2}
\right) dr^2 + t^2 \sum_{i=1}^3\left(dx^i\right)^2\ .
\ee
This spacetime can be regarded as a cosmological one, which has a singularity 
at $t=0$, which may be identified with the big-bang singularity. When 
$t\to \infty$, the spacetime approaches to the deSitter spacetime. If we
put a 
brane at $t\to\infty$, there will exist a dual conformal field theory
(dS/CFT correspondence \cite{strominger,ds1}). By repeating 
a similar calculation as the asymptotically anti-deSitter case after 
replacing the $l^2$ by $-l^2$, we can evaluate the conformal anomaly at 
$t\to\infty$ and obtain the expression identical with (\ref{II11}).
This finishes our proof of independence of holographic conformal anomaly 
from the bulk space choice. However, it could be that in different dualities 
(say AdS or dS) the corresponding dual CFTs having same central charges
could be essentially different.

\section{Newton potential of the gravity induced on the brane 
in the bulk AdS black hole spacetime}

In the present section our problem is to describe  the 
calculation of Newton potential of the gravity localized on the 
3-brane in the 5 dimensional Schwarzschild-AdS background. 
Metric is chosen  in the warped form:
\be
\label{N1}
ds^2=dy^2 + \e^{2A(y)}\sum_{\mu,\nu=0}^3g_{\mu\nu}dx^\mu dx^\nu \ ,
\ee
the action of the brane at $y=y_0$ is given by
\be
\label{N2}
S_b={2 \kappa_0 \over \kappa^2} \delta\left(y-y_0\right)
\int d^4 x \sqrt{-g}\ .
\ee
The metric (\ref{N1}) can be rewritten in the Schwarzschild-like 
form:
\be
\label{N3}
ds^2 = \sum_{m,n=0}^4 G_{mn}dx^m dx^n
=\e^{-2\rho(r)}dr^2  - \e^{2\rho}dt^2 + r^2 
\sum_{i,j=1}^3 \tilde g_{ij}dx^i dx^j\ .
\ee
Here 
\be
\label{N4}
dy=\e^{-\rho}dr\ .
\ee
Then since $dy\delta\left(y -y_0\right)=dr \delta
\left(r-r_0\right)$ ($r_0=r\left(y_0\right)$), 
the brane action (\ref{N2}) is rewritten as 
\be
\label{N5}
S_b={2 \kappa_0 \over \kappa^2} 
\delta\left(r-r_0\right)\e^{-\rho\left(r_0\right)}
\int d^4x \sqrt{ -g}\ .
\ee
As a solution of the vacuum Einstein equation, we consider 
the Schwarzschild AdS metric (\ref{N3}) 
\be
\label{N6}
\e^{2\rho}={r^2 \over l^2} - {\mu \over r^2}\ ,\quad 
\sum_{i,j=1}^3 \tilde g_{ij}dx^i dx^j
=\sum_{i,j=1}^3 \left(dx^i\right)^2 \ .
\ee
Here we assume that the shape of the event horizon is 
flat. If we orbifoldize the spacetime by identifying 
$r-r_0=-\left(r - r_0\right)$, the dynamics of the brane 
is described by the FRW like equation:
\be
\label{N7}
H^2= - {1 \over l^2} + {\kappa_0^2 \over 4} + {\mu \over a^4}\ .
\ee
Here $l^2$ is the length parameter, which is related with 
the bulk cosmological constant $\Lambda$ by $\Lambda
= - {12 \over l^2}$. We also assume that there is a brane 
at $r=a(\tau)$ ($\tau$ is 
the proper time on the brane) and the Hubble parameter $H$ 
is defined by $H\equiv {1 \over a}{da \over d\tau}$. 
Then if the brane is static ($H=0$), we obtain
\be
\label{N8}
a^4=r_0^4= {\mu \over {1 \over l^2} - {\kappa_0^2 \over 4}}\ .
\ee
On the other hand, with the metric assumption (\ref{N3}), 
the $(i,j)$-component of the bulk Einstein equation gives
\bea
\label{N9}
&& -{2 \over r^2}\left(1 +\rho' r\right)\e^{2\rho} 
 - {1 \over 2}\e^{2\rho}\left\{-2\rho'' - 4\left(\rho'\right)^2 
 - {12 \over r} \rho' - {6 \over r^2}\right\} 
 + {\Lambda \over 2} \nn
&=& \kappa_0 \e^{-\rho\left(r_0\right)}
 \delta\left(r-r_0\right)
\eea
and 
\be
\label{N10}
\left.\left(\e^{2\rho}\right)''\right|_{r=r_0}=2
\kappa_0 \e^{-\rho\left(r_0\right)}\delta\left(r-r_0\right)\ .
\ee
The above equation (\ref{N10}) is rewritten as 
\be
\label{N11}
\kappa_0=-{2 \over l}\left(1 + {\mu l^2 \over r_0^4}\right)
\left(1 - {\mu l^2 \over r_0^4}\right)^{1 \over 2}\ .
\ee
By combining (\ref{N8}) and (\ref{N11}), one obtains
\be
\label{N12}
\alpha^3 + \alpha^2 - 3\alpha=0\ ,\quad
\alpha\equiv {\mu l^2 \over r_0^4}. 
\ee
Then the brane can exist at 
\be
\label{N13}
\alpha= {\mu l^2 \over r_0^4}={-1 + \sqrt{13} \over 2}\ .
\ee
We should note, however, the brane is not stable. In fact, 
if we rewrite the FRW equation (\ref{N7}) as
\be
\label{N14}
\left({da \over d\tau}\right)^2=-V(a)\ ,\quad 
V(a)=-\left({\kappa_0^2 \over 4} - {1 \over l^2}\right)a^2 
- {\mu \over a^2}\ ,
\ee
we find the effective potential has no minimum 
since  assumption ${\kappa_0^2 \over 4} - {1 \over l^2}<0$ in 
order that static brane exists, the effective potential $V(a)$ is 
monotonically increasing function. Then the brane is attracted by 
the black hole and falls into the black hole. 

In the following we assume, for simplicity, that the brane is 
adiabatically static by considering the behavior around the 
case of (\ref{N8}), where the brane can be static for a 
short time interval, or by introducing the another force like 
electromagnetic interaction, which might stabilize the brane. 

The Einstein equation is given by
\be 
\label{N15}
R_{mn} - {1 \over 2}G_{mn} R + {\kappa^2\Lambda \over 2}G_{mn}
=-2\left({\mu \over r_0^3} + {r_0 \over l^2}\right)
\delta\left(r-r_0\right)G_{mn}\ .
\ee
Then if we consider the perturbation of the metric 
$G_{mn}\to G_{mn} + \delta G_{mn}$, the Einstein equation 
(\ref{N15}) gives
\bea
\label{N16}
0&=& \left.{1 \over 2}\right\{\nabla^l\nabla_m \delta G_{nl} 
+ \nabla^l\nabla_n \delta G_{ml} - \nabla^2 \delta G_{mn} \nn
&& \left.  - {1 \over 2}\left(\nabla_m\nabla_n + \nabla_n\nabla_m\right)
 \left(G^{kl}\delta G_{kl}\right)\right\} \nn
&& - {1 \over 2}\delta G_{mn} \left( R - \kappa^2 \Lambda \right)
+ 4\kappa^2\left({\mu \over r_0^3} + {r_0 \over l^2}\right)
\delta G_{mn}\delta\left(r-r_0\right) \nn
&& - {1 \over 2}G_{mn}\left( - \delta G_{kl} R^{kl} 
 - \nabla^k \nabla^l \delta G_{kl} 
 - \nabla^2 \left( G^{kl}\delta G_{kl}\right) \right)\ .
\eea
Here the curvature $R$ and $R_{mn}$, the covariant derivative 
$\nabla_m$ are defined by the unperturbative part $G_{mn}$ 
of the metric. We now choose the following gauge conditions
\be
\label{N17}
\nabla^m \delta G_{mn}= \delta G_{rm} = \delta G_{mr} 
= \delta G_{tm} = \delta G_{mt} = 0\ ,
\ee
and we write $\delta G_{ij} = h_{ij}$. Then by using the 
solution (\ref{N6}) of the bulk Einstein equation, 
the $(m,n)=(i,j)$ components of Eq.(\ref{N16}) are given 
by
\bea
\label{N18}
0&=& -{1 \over 2}\left[{1 \over r^2}\triangle h_{ij}
 - \left({r^2 \over l^2} - {\mu \over r^2}\right)^{-1}
\partial_t^2 h_{ij} 
+ \left({r^2 \over l^2} - {\mu \over r^2}\right)\left\{
\partial_r^2 h_{ij} - {1 \over r}\partial_r h_{ij} 
\right.\right.\nn 
&& \left.\left. +2\left({r \over l^2} + {\mu \over r^3}\right) 
\left(\partial_r h_{ij} - {2 \over r}h_{ij}\right)\right\}
\right] - 2 \kappa^2\left({\mu \over r_0^3} + {r_0 \over l^2}\right)
\delta\left(r-r_0\right) h_{ij} \ ,\nn
\triangle&\equiv&\sum_{k=1}^3\left(\partial_k\right)^2\ .
\eea
Taking the plane wave on the brane as
\be
\label{N19}
h_{ij}=h^{(0)}_{ij}\e^{i\sum_{l=1}^3 k_lx^l - i\omega t}
\left({r \over l} - {\mu l \over r^3}\right)^{-{1 \over 2}}
\phi(r) \ ,
\ee
one gets
\bea
\label{N20}
0&=& - {1 \over 2}\left[ \partial_r^2  
+ {1 \over r^2}\left\{ - {15 \over 4} + {4\mu^2 \over r^4}
\left({r^2 \over l^2} - {\mu \over r^2}\right)^{-2}\right\} 
\right.\nn
&& - \left({r_0^2 \over l^2} - {\mu \over r_0^2}\right)^{-1}
\left({3r_0 \over l^2} + {\mu \over r_0^3}\right)\delta
\left(r-r_0\right) \nn
&& \left. 
 - \left\{\left({r^2 \over l^2} - {\mu \over r^2}\right)^{-1}
{k^2 \over r^2} - \left({r^2 \over l^2} 
 - {\mu \over r^2}\right)^{-2}\omega^2\right\}\right]\phi\ .
\eea
When $\mu=0$, the above equation reduces to that in the 
Randall-Sundrum model\cite{RS2}: 
\bea
\label{N20b}
0&=& - {1 \over 2}\left[ \partial_r^2  
 - {15 \over 4r^2} - {3 \over r_0} \delta
\left(r-r_0\right) - {l^4 \over r^4}
\left({k^2 \over l^2} - \omega^2\right)\right]\phi\ .
\eea
In fact, by identifying $r=r_0\e^{-{|y| \over l}}$ and 
redefining $\phi$ as $\phi=\e^{-{|y| \over 2l}}\tilde \phi$, 
Eq.(\ref{N20b}) can be rewritten as
\bea
\label{N20c}
0&=& - {1 \over 2}\left[ \partial_y^2  
 - {4 \over l^2} - {2 \over l} \delta
\left(y\right) + m^2\e^{2|y| \over l} \right]
\tilde \phi\ ,\quad 
m^2\equiv -{l^2 \over r_0^2}
\left({k^2 \over l^2} - \omega^2\right)\ ,
\eea
which is identical with the corresponding equation (8) 
in \cite{RS2}. 

For small $\mu$, the graviton 
will be localized on the brane. We should note, however, the 
condition that the graviton is massless on the brane is given 
by
\be
\label{N21}
-m^2\equiv {k^2 \over r_0^2} - \left({r_0^2 \over l^2} 
 - {\mu \over r_0^2}\right)^{-1}\omega^2 =0\ .
\ee
Then not as in the case of the Randall-Sundrum model, the 
equation (\ref{N20}) does not reduce to the eigenvalue 
equation.

The radius $r_H$ of the event horizon is given by
\be
\label{N22}
r_H=\mu^{1 \over 4}l^{1 \over 2}\ .
\ee
We now consider the large black hole ${\mu \over l^2}
\to \infty$ or ${r_H \over l}\to \infty$ by fixing 
$r_0-r_H$. 
Then by defining a new coordinate $\xi$ as
\be
\label{N24}
r=r_H + \xi\ ,
\ee
Eq.(\ref{N20}) reduces into 
\be
\label{N25}
0=\partial_\xi^2\phi - {15 \over 4 r_H^2} - {l^2k^2 \over 4 r_H^3 \xi} 
+ \left({1 \over 4 } + {l^4 \omega^2 \over 16 r_H^2}\right)
{1 \over \xi^2}\phi 
 - {1 \over \xi_0}\delta\left(\xi - \xi_0\right)\phi\ .
\ee
Here we assume that there is a brane at $\xi=\xi_0
\equiv r_0 - r_H$. 
When $k^2=0$, we can solve Eq.(\ref{N25}) by
\be
\label{N26}
\phi = \left({\xi \over l}\right)^{1 \over 2}
\left(\alpha I_{i{l^2\omega \over 4r_H}}\left({\sqrt{15} \over 2r_H}\xi\right)
+ \alpha^* I_{-i{l^2\omega \over 4r_H}}\left({\sqrt{15} \over
2r_H}\xi\right)\right)\ .
\ee
Here $\alpha$ is a complex constant and $I_\nu(z)$ is a deformed Bessel
function. 
Since we are now considering the large black hole, we may approximate the
modified 
Bessel functions by
\be
\label{N27}
I_\nu(z)\sim {1 \over \Gamma \left(1+\nu\right)}\left({z \over 2}\right)^\nu 
+ \cdots \ .
\ee
Then the phase of $\alpha$ can be determined by requiring that the solution
satisfies 
the delta function in (\ref{N25}), that is, by the equation;
\be
\label{N28}
 -2\left.{\partial_\xi \phi \over \phi}\right|_{\xi=\xi_0}={1 \over \xi_0}\ ,
\ee
which leads
\be
\label{N29}
-{\alpha^* \over \alpha}
= \left({\sqrt{15} \over 2r_H}\xi_0\right)^{2i{l^2\omega \over 4r_H}}
{\Gamma\left(1 - i{l^2\omega \over 4r_H}\right) \over 
{\Gamma\left(1 + i{l^2\omega \over 4r_H}\right)}}
{1 + i{l^2\omega \over 4r_H} \over 1 - i{l\omega \over 4r_H}}\ .
\ee
There are some ambiguities how one should treat 
the boundary condition of $\phi$ at the horizon. 
In order to avoid this problem, we put one more brane with tension 
$-{1 \over l}$ at $\xi=\xi_1<\xi_0$ as in the first Randall-Sundrum model. 
Then in addition to Eq.(\ref{N29}), we obtain another condition;
\be
\label{N30}
-{\alpha^* \over \alpha}
= \left({\sqrt{15} \over 2r_H}\xi_1\right)^{2i{l^2\omega \over 4r_H}}
{\Gamma\left(1 - i{l^2\omega \over 4r_H}\right) \over 
\Gamma\left(1 + i{l^2\omega \over 4r_H}\right)}
{1 + i{l^2\omega \over 4r_H} \over 1 - i{l^2\omega \over 4r_H}}\ .
\ee
Combining (\ref{N29}) and (\ref{N30}) leads to
\be
\label{N30a}
1=\left({\xi_0 \over \xi_1}\right)^{2i{l^2\omega \over 4r_H}}
=\e^{2i{l^2\omega \over 4r_H}\ln \left({\xi_0 \over \xi_1}\right)}\ .
\ee
That is
\be
\label{N31}
{l\omega \over 4r_H}\ln \left({\xi_0 \over \xi_1}\right)=\pi n \ ,
\quad n=0,1,2,3,\cdots\ .
\ee
Then by using (\ref{N21}), we find 
\be
\label{N32}
m^2={4r_H \over \xi_0l^2}\left({\pi n \over \ln \left({\xi_0 \over
\xi_1}\right)}\right)^2\ .
\ee
Then for finite $\xi_1$, the Kaluza-Klein modes, which correspond to
$n=1,2,3,\cdots$, 
become very heavy since we are now considering the big black hole, where
$r_H$ is large. 
Then the Kaluza-Klein modes  decouple on the brane and only massless mode, 
corresponding to $m^2=0$ or $n=0$, will contribute to the Newton potential. 
Therefore if $r$ is the distance between two particles on the brane, the 
Newton potential  behaves as ${1 \over r}$ and the graviry would localize on 
the brane for the two brane model. We should note that such a dcoupling of
the 
Kaluza-Klein modes makes the Newton potential on  both  branes 
corresponding to $\xi_0$ and $\xi_1$ to behave as ${1 \over r}$. 
If the inner brane, which exists 
at $\xi=\xi_1$, approaches to the horizon $\xi_1\to 0$, however, 
the logarithtic term in (\ref{N30a}) becomes dominant and the Kaluza-Klein 
modes become light. Then the Newton potential  behaves as ${1 \over r^2}$, 
as in the five dimensional one. This may indicate that, for one brane
model, the gravity 
would not localize on the brane. 

 Thus, we demonstrated that at sufficiently reasonable conditions 
the RS2 (two branes) model realized in bulk AdS BH shows the properties
similar to the properties of RS2 model in the pure AdS bulk \cite{RS2}.
Specifically, gravity trapping on the brane occurs. 
Note that gravity trapping on the brane occurs also when bulk is 5d de
Sitter space \cite{dstrap}. In this sense one can say again about (bulk
metric) independence 
of the brane gravity trapping from the choice of bulk spacetime.
Moreover, the fact that bulk space is AdS BH is not essential too.
In other words, again the localization is universal feature of holographic
duality
(no horizon dependence).

\section*{Acknowledgments}

The work by S.N. is supported in part by the Ministry of Education, 
Science, Sports and Culture of Japan under the grant n. 13135208.


\begin{thebibliography}{99}
\bibitem{AdS} J.M. Maldacena, {\sl Adv.Theor.Math.Phys.} {\bf 2} (1998) 231;
E. Witten, {\sl Adv.Theor.Math.Phys.} {\bf 2} (1998) 253;
S. Gubser, I. Klebanov and A. Polyakov, {\sl Phys.Lett.} 
{\bf B428} (1998) 105, 
O. Aharony, S. Gubser, J. Maldacena, H. Ooguri and Y. Oz,
{\it Phys. Rept. \/} {\bf 323} 183, (2000), hep-th/9905111.
\bibitem{anom} M. Henningson and K. Skenderis, 
{\sl JHEP} {\bf 9807} (1998) 023;  S. Nojiri and S. D. Odintsov, 
{\sl Phys.Lett.} {\bf B444} (1998) 92, hep-th/9810008;
 M. Nishimura and Y.
Tanii, 
{\sl Int.J.Mod.Phys.} {\bf A14} (1999) 3731; 
V. Balasubramanian and P. Kraus, 
{\sl Commun.Math.Phys.} {\bf 208} (1999) 3731; 
S. Nojiri and S.D. Odintsov, 
{\sl Int.J.Mod.Phys.} {\bf A15} (2000) 413, hep-th/9903033; 
{\sl Mod.Phys.Lett.} {\bf A15} (2000) 1043, hep-th/9910113; 
M. Blau,K. Narain and E. Gava, {\sl JHEP} {\bf 9909} (1999) 018;
A. Bilal and C.-S. Chu, {\sl Nucl.Phys.} {\bf B562} (1999) 181,
hep-th/9907106; 
W. Muck and K.S. Wiswanathan, hep-th/9905046; 
P. Mansfield and D. Nolland, 
{\sl JHEP} {\bf 9907} (1999) 028; hep-th/0005224; 
J. Ho, hep-th/9910124; 
C. Imbimbo, A. Schwimmer, S. Theisen and S. Yankielowicz,
hep-th/9910267; 
F. Bastianelli, S. Frolov and A. Tseytlin, hep-th/0001041;
S. Nojiri,S.D. Odintsov and S. Ogushi, {\sl Phys.Rev.} {\bf D62}
(2000) 124002, hep-th/0001122; {\sl Phys.Lett.} {\bf B494} (2000) 318,
hep-th/0009015; hep-th/0011182;
M. Fukuma, S. Matsuura and T. Sakai, hep-th/0007062; hep-th/0103187;
hep-th/0204257;
G. Naculich, H.J. Schnitzer and N. Wyllard, hep-th/0106020;
A.M. Ghezelbash and R. Mann, hep-th/0210046.
\bibitem{dsanom} S. Nojiri and S.D. Odintsov, hep-th/0106191,
{\sl Phys.Lett.} {\bf B519} (2001) 145.
\bibitem{strominger} A. Strominger, hep-th/0106113.
\bibitem{ds1} C.M. Hull, {\sl JHEP} {\bf 9807} (1998) 021, hep-th/9806146.
\bibitem{RS2} L. Randall and R. Sundrum,
{\sl Phys.Rev.Lett.} {\bf 83} 3370 (1999), hep-th/9905221; 
{\sl Phys.Rev.Lett.}  {\bf 83} 4690 (1999), hep-th/9906064.
\bibitem{dstrap} S. Nojiri and S.D. Odintsov, hep-th/0107134,
{\sl JHEP} {\bf 0112} (2001) 033;
I. Brevik, K. Ghoroku, S.D. Odintsov and M. Yahiro, hep-th/0204066.
\end{thebibliography}
\end{document}